\documentclass[twocolumn,PRD,showpacs,nofootinbib,amsmath,amssymb]{revtex4}

\usepackage{graphicx}
\usepackage{dcolumn}
\usepackage{bm}

\def\veck{\mathbf{k}}
\def\hatn{\mathbf{\hat n}}
\def\rmX{\mathrm{X}}
\def\rmT{\mathrm{T}}
\def\rmE{\mathrm{E}}
\def\rmA{\mathrm{A}}
\def\rmn{\mathrm{n}}
\def\VEV#1{\left\langle #1 \right\rangle}
\begin{document}

\title{Cosmic Microwave Background Statistics for a
Direction-Dependent Primordial Power Spectrum}

\author{Anthony R. Pullen and Marc Kamionkowski}
\affiliation{California Institute of Technology, Mail Code 130-33,
Pasadena, CA 91125}

\date{\today}

\begin{abstract}
Statistical isotropy of primordial perturbations is a common
assumption in cosmology, but it is an assumption that should be
tested.  To this end, we develop cosmic microwave background
statistics for a primordial power spectrum that depends on the
direction, as well as the magnitude, of the Fourier wavevector.
We first consider a simple estimator that searches in a
model-independent way for anisotropy in the square of the
temperature (and/or polarization) fluctuation.  We
then construct the minimum-variance estimators for the
coefficients of a spherical-harmonic expansion of the
direction-dependence of the primordial power spectrum.  To illustrate, we
apply these statistics to an inflation model with a quadrupole
dependence of the primordial power spectrum on direction and
find that a power quadrupole as small as 2.0\% 
can be detected with the Planck satellite.
\end{abstract}


\maketitle

\section{Introduction} \label{S:intro}

It is well known that the homogeneity and isotropy of the
Universe are only approximate.  There are departures from
homogeneity and isotropy that are now well-quantified by
measurements of the cosmic microwave background (CMB) and galaxy
surveys.  In current cosmological theory, the notions of
homogeneity and isotropy have been superseded by the notions of
{\it statistical} homogeneity and isotropy.  The density of matter
may differ from one point in the Universe to another, but the
distribution of matter is described as a realization of a random
field with a variance that is everywhere the same and
the same in every direction.  This is generally the
prediction of structure-formation models, and in particular, of
inflationary models.

Still, statistical isotropy and homogeneity are assumptions that
can be tested quantitatively, and the precision with which they
can be tested is improving rapidly with the still-accumulating
wealth of cosmological data.  Preliminary (and controversial)
indications for a preferred direction in the CMB
\cite{preferreddirections,Eriksen:2003db,Hansen:2004vq,Eriksen:2007pc}
have recently motivated the study of departures from statistical 
isotropy.  Subsequent theoretical work has shown that
although statistical isotropy is a generic prediction of
inflation, inflation models can in fact be constructed to
violate statistical isotropy
\cite{Chibisov:1989wb,Chibisov:1990bk,Berera:2003tf,Donoghue:2004gu,Buniy:2005qm,Ackerman:2007nb,Gumrukcuoglu:2007bx,ArmendarizPicon:2007nr,Donoghue:2007ze,Pereira:2007yy}.
Dark-energy models might also accommodate departures
from statistical isotropy \cite{Battye:2006mb,Koivisto:2007bp}.  These
models provide useful straw men against which
the success of the standard inflationary predictions of
statistical isotropy can be quantified.

The growing interest in such models motivates us to study
generalized tests for statistical isotropy.
In a statistically isotropic Universe, the primordial distribution of
matter is a realization of a random field in which Fourier modes
of the density field have variances, a power spectrum
$P(\veck)$, that depend only on the magnitude $k$ of the
wavevector $\veck$.  If we drop the assumption of statistical
isotropy, the power spectrum will depend on the direction
$\mathbf{\hat k}$ as well.  If $\delta(\veck)$ is the Fourier
amplitude of the fractional density perturbation, then the power
spectrum is defined by
\begin{equation}
     \VEV{\delta(\veck)\delta^*(\veck')} = 
     \delta_D(\veck-\veck')P(\veck),
\end{equation}
where the angle brackets denote an average over all realizations
of the random field, and $\delta_D$ is a Dirac delta function;
note that we are still preserving the assumption that different
Fourier modes are uncorrelated.  The most general power spectrum
can then be written,
\begin{equation}   \label{E:powerspectrum}
     P(\veck)= A(k) 
     \left[1+ \sum_{LM} g_{LM}(k) Y_{LM}(\mathbf{\hat{k}}) \right],
\end{equation}
where $Y_{LM}(\mathbf{\hat k})$ (with $L\geq2$) are spherical
harmonics, and $g_{LM}(k)$
quantify the departure from statistical isotropy as a function of
wavenumber $k$.  Since the density field is real, Fourier modes
for $\veck$ are related to those of $-\veck$, in such a way that
the multipole moment $L$ must be even.  In the limit
$g_{LM}(k)\rightarrow0$, we recover the usual statistically
isotropic theory with power spectrum $A(k)$.  The
implementation, Eq.~(\ref{E:powerspectrum}), of power anisotropy
is motivated in part by the inflationary model of
Ref.~\cite{Ackerman:2007nb}, which predicts $g_{2M}(k)\neq0$.

Here we consider several CMB tests for statistical
isotropy.  The first, which we refer to as ``power multipole
moments,'' is a simple and intuitive estimator that involves
measurement of the multipole moments of the square of the
temperature/polarization fields.\footnote{There
has already been
some evidence for a dipole in the CMB power
\cite{Eriksen:2003db,Eriksen:2007pc} that is analogous to the
higher multipole moments that we are considering here, but which cannot be due
to anisotropy in the primordial power spectrum because it has $L=1$.
There have also been searches \cite{Hansen:2004vq} for
anisotropy along the lines considered here, and
Ref.~\cite{Hajian:2003qq,ArmendarizPicon:2005jh} discusses
similar statistics.}
As an example, we apply this statistic to an inflationary model
\cite{Ackerman:2007nb} that predicts a quadrupole in the matter
power spectrum.  

Although power multipole moments provide a nice
model-independent test for departures from statistical
isotropy, more sensitive probes can be developed if the
particular form of the departure is specified.  
To illustrate, we thus construct the minimum-variance estimators for
the anisotropy coefficients $g_{LM}(k)$ under the assumption that
they are constants.  The naive power
multipole moments, although intuitively simple, co-add a number
of modes with equal weight.  The minimum-variance estimator
co-adds modes with weights that depend on their signal to
noise, so that (as the name suggests) the variance of the
estimator is minimized.  We show that this statistic provides a far
stronger probe for the $g_{LM}$s.

The plan of our paper is as follows: Section \ref{S:prelim}
reviews some CMB basics.  Section \ref{S:offdiagonal}
calculates the correlations of CMB spherical-harmonic coefficients
if there are departures from statistical isotropy.  As we
discuss there, the power spectrum $C_l$, which describes the
two-point CMB statistics if there is statistical isotropy, is
generalized to a set of moments $D^{LM}_{ll'}$ if statistical
isotropy is broken.  In Section
\ref{S:powermoments}, we introduce and calculate the power
multipole moments and calculate the standard errors with which
these moments can be recovered.  We apply this statistic to a
quadrupole in the matter power spectrum, calculating the
sensitivities of several CMB experiments to
such a quadrupole.  Section \ref{S:optimal} discusses
minimum-variance estimators for the quantities $D^{LM}_{ll'}$
that parametrize the 
departures $g_{LM}(k)$ from statistical isotropy.  We then
construct from these the 
minimum-variance estimators for the quadrupole moments of the
primordial power spectrum, calculate their variance, and
evaluate their sensitivity to departures from statistical
isotropy.  We make some concluding remarks in
Section~\ref{S:conclusions}.  Throughout the main body of the
paper, we discuss statistics for only a temperature map, in
order to make the presentation clear.  An Appendix generalizes
to include the full temperature-polarization information.
Our numerical results are for a full temperature-polarization
map, as well as for temperature or polarization alone.

\section{Preliminaries} \label{S:prelim}

A CMB experiment provides the temperature $T(\hatn)$ 
as a function of position $\hatn$ on the sky.  
The map $T(\hatn)$ can be expanded in terms of 
spherical harmonics $Y_{lm}(\hatn)$,
\begin{equation} \label{E:alm}
     a_{lm} = \frac{1}{T_0}\int
     d\hatn \,Y_{lm}^\ast(\mathbf{\hat{n}})T(\hatn).
\end{equation}
The $a_{lm}$s are Gaussian random variables, and if there is
statistical isotropy, then they are statistically independent
for different $l$ and $m$: $\VEV{a_{lm} a_{l'm'}^*} =
C_l \delta_{ll'}\delta_{mm'}$.\footnote{Strictly speaking, it is
not the $a_{lm}$s that are statistically independent, but rather
their real and imaginary parts.}  The set of $C_l$s is the CMB
temperature power spectrum.  We will see that
when statistical isotropy is violated, there are correlations
induced between $a_{lm}$s for different $l$ and $m$
\cite{Ackerman:2007nb}.  If there is statistical isotropy, the
two-point autocorrelation function is
\begin{eqnarray} \label{E:corrfn}
     C(\hatn,\hatn^\prime) &=&
     \VEV{T(\hatn)T(\hatn')} \nonumber \\
     &=&
     T_0^2\sum_l\frac{2l+1}{4\pi} C_l P_l(\hatn\cdot\hatn');
\end{eqnarray}
i.e., the correlation function depends only on the separation
between the two points.  If statistical isotropy is violated,
this is not necessarily true.

\section{Off-diagonal correlations for anisotropic power}
\label{S:offdiagonal}

Consider a primordial matter power spectrum $P(\mathbf{k})$
given by Eq.~(\ref{E:powerspectrum}).
We expand $T(\mathbf{\hat{n}})$ in $\veck$
space in the form,
\begin{equation} \label{E:temp}
     \frac{T}{T_0}(\mathbf{\hat{n}}) = \int
     d^3k\,\sum_l(-i)^l(2l+1)
     P_l(\mathbf{\hat{k}}\cdot\mathbf{\hat{n}})
     \delta(\mathbf{k})\Theta_l(k),
\end{equation}
where $\Theta_l(k)$ is the contribution to the $l$th temperature
moment from wavevector $\veck$.  With these conventions,
$\Theta_l(k)$ is real.  With our expression,
Eq.~(\ref{E:powerspectrum}), we can write the covariance
matrix as
\begin{equation} \label{E:cov}
     \VEV{ a_{lm} a_{l^\prime m^\prime}^* }
     = \delta_{ll^\prime}\delta_{mm^\prime} C_l
     + \sum_{LM} \xi^{LM}_{lml^\prime  m^\prime} D^{LM}_{ll^\prime}.
\end{equation}
Here, the set of $C_l$s, given by
\begin{equation}  \label{E:cls}
  C_l = (4\pi)^2\int_0^\infty
 dk\,k^2A(k)[\Theta_l(k)]^2,
\end{equation}
is the usual CMB power spectrum for the case of statistical
isotropy.  Departures from statistical
isotropy introduce the second term, where
\begin{equation} \label{E:cldl}
     D^{LM}_{ll^\prime} = (4\pi)^2(-i)^{l-l^\prime}\int_0^\infty
     dk\, k^2 A(k) g_{LM}(k) \Theta_l(k)\Theta_{l^\prime}(k),
\end{equation}
and
\begin{eqnarray} \label{E:xint}
     \xi^{LM}_{lml^\prime m^\prime} &=&
     \int d\mathbf{\hat{k}} \, Y_{lm}^\ast(\mathbf{\hat{k}})
     Y_{l^\prime m^\prime}(\mathbf{\hat{k}})
     Y_{LM}(\mathbf{\hat{k}}) \nonumber \\
     &=& (-1)^{m'} \left(G^L_{ll'} \right)^{1/2}
     C^{LM}_{lml',-m'},\nonumber \\
\end{eqnarray}
where $C_{lml^\prime m^\prime}^{LM}$ are Clebsch-Gordan
coefficients, and
\begin{equation}
     G^L_{ll'} \equiv \frac{ (2l+1)(2l'+1)}{ 4 \pi (2L+1) }
     \left(C^{L0}_{l0l'0} \right)^2.
\end{equation}
Throughout, we use upper-case indices $LM$ for
power anisotropies, and lower-case indices $lm$ for
temperature/polarization anisotropies.  For $L$ even,
$\xi_{lml'm'}$ are nonvanishing only for $l-l'$ even, and so the
$D^{LM}_{ll'}$ are real.  Eqs.~(\ref{E:cldl}) and
(\ref{E:xint}) agree with similar results in
Ref.~\cite{ArmendarizPicon:2005jh}, and  they recover the
results of Ref.~\cite{Ackerman:2007nb} for $L=2$.

If primordial perturbations are statistically isotropic and
Gaussian, then the statistics of the CMB temperature map are
specified fully by the power spectrum, the set of $C_l$s.  If
primordial perturbations have a departure from statistical
isotropy that can be written in terms of spherical harmonics
$Y_{LM}(\mathbf{\hat k})$, then the two-point statistics are
described additionally by the set of multipole moments
$D^{LM}_{ll'}$.  These quantities are thus the generalization
of the $C_l$s if there is statistical anisotropy.

\section{Power multipole moments} \label{S:powermoments}

\subsection{Theoretical predictions}

It is natural to expect that a spherical-harmonic pattern of
anisotropy in the matter power spectrum manifests itself in a
similar pattern in the CMB power.  It is thus natural to
consider a set of ``power multipole moments,''
\begin{equation} \label{E:blm}
     b_{LM} = \frac{1}{T_0^2} \int
     d\hatn\,Y_{LM}^\ast(\mathbf{\hat{n}}) \VEV{T^2}(\hatn),
\end{equation}
where $\VEV{T^2}(\hatn)=C(\hatn,\hatn)$ is the
expectation value of the square of the temperature at position
$\hatn$ in the sky; it is the autocorrelation
function at zero lag.  With this statistic, we
simply look for anisotropies in the power.  These statistics
have several advantages.  In addition to having a form familiar
from similar statistics [e.g., Eq.~(\ref{E:alm})] for
temperature fluctuations, they have
simple analytic expressions in terms of
$P(\veck)$.  There are also (as we show below), relatively
simple expressions for the cosmic-variance-- and
instrumental-noise--induced errors in the measurement of these
statistics.  

The variance $\VEV{T^2}(\hatn)$ as a function of position
$\hatn$ is given by
\begin{equation} \label{E:cn}
     \frac{\VEV{T^2}(\hatn)}{T_0^2} = \sum_{lml^\prime
     m^\prime}\VEV{ a_{lm} a_{l^\prime
     m^\prime}^{\ast} }
     Y_{lm}(\mathbf{\hat{n}})Y_{l^\prime
     m^\prime}^\ast(\mathbf{\hat{n}}).
\end{equation}
We put this into Eq.~(\ref{E:blm}) and use
\begin{equation} \label{E:CGortho}
     \sum_{mm'} C^{LM}_{lml,-m'} C^{L'M'}_{lml',-m'} =\delta_{LL'}
     \delta_{MM'},
\end{equation}
to obtain (for $L\geq2$)
\begin{equation} \label{E:blmtot}
     b_{LM} = 
     \sum_{l l'} G^L_{ll'}D^{LM}_{ll'}.
\end{equation}

\subsection{Statistical noise}
\label{S:unips2}

We now calculate the standard error, due to cosmic variance and
instrumental noise, with which the power multipole moments can
be measured.  To do so, we consider a full-sky map
$T^{\rm{map}}(\mathbf{\hat{n}})$ of the temperature
in $N_{\rm{pix}}$ equal-area
pixels.  The temperature in each pixel receives
contributions from signal and from noise.  Thus, in pixel $i$,
$T^{\rm map}=T(\hatn_i) + T^\rmn_i$, where $T(\hatn_i)$ is the
temperature measured in pixel $i$, which will be the signal
temperature smoothed by a Gaussian beam of full-width half
maximum (fwhm) $\theta_{\rm fwhm}$, plus a noise $T_i^\rmn$.
We assume that the noise is isotropic and that the
noises in different pixels are uncorrelated
with variance $\sigma_T^2$: i.e., $\VEV{T_i^\rmn T_j^\rmn}=\sigma_T^2
\delta_{ij}$.  The power spectrum for the map is thus
$C_l^{\rm{map}} = |W_l|^2C_l+ C_l^\rmn$, where $C_l^{\rmn} =
(4\pi/N_{\rm{pix}})\sigma_T^2$ is the noise power spectrum, and
$W_l$ is a window function that takes into account the effects
of beam smearing; for a Gaussian beam of fwhm $\theta_{\rm
fwhm}$, it is $W_l = \exp(-l^2\sigma_b^2/2)$ with $\sigma_b =
\theta_{\rm{fwhm}} / \sqrt{8\ln2} =
0.00742(\theta_{\rm{fwhm}}/1^\circ)$.

Since the instrumental noise is isotropic by assumption, we
get an unbiased estimator for $b_{LM}$ (for $L\geq2$)
from
\begin{equation} \label{E:dest}
     \widehat{b}_{LM}^{\rm map} = \frac{1}{T_0^2}\int
     d\hatn\,Y_{LM}^\ast(\mathbf{\hat{n}})
     \left[T^{\rm{map}}(\mathbf{\hat{n}})\right]^2.
\end{equation}
Cosmic variance and instrumental noise induce a variance in the
$b_{LM}$s, which we define as
\begin{equation} \label{E:covmat2}
     \Xi^{LM}  \equiv  \VEV{
     \widehat{b}_{LM}\widehat{b}_{LM} },
\end{equation}
where we have assumed the null hypothesis, $g_{LM}=0$.  For this
null hypothesis of a statistically isotropic Gaussian random map, 
\begin{eqnarray} \label{E:xlm}
     \Xi^{LM} &=& \frac{2}{T_0^4}\int
     d\hatn\,d\hatn'\,
     C^{\rm{map}}(\mathbf{\hat{n}},\mathbf{\hat{n}^\prime})
     C^{\rm{map}}(\mathbf{\hat{n}},\mathbf{\hat{n}^\prime})
     \nonumber \\ 
     & & \times Y_{LM}(\mathbf{\hat{n}})Y_{LM}^\ast(\mathbf{\hat{n}^\prime})
     \nonumber      \\ 
     &=& 2 \sum_{ll'} G^L_{ll'} C_{l}^{\rm{map}} 
     C_{l'}^{\rm{map}} \nonumber \\
\end{eqnarray}
where $C^{\rm{map}}
(\mathbf{\hat{n}}_1,\mathbf{\hat{n}}_2)$ is the two-point 
correlation function for the {\it map}, obtained from the expression,
Eq.~(\ref{E:corrfn}), for the correlation function by replacing
$C_l$ by $C_l^{\rm map}$, and we have used $\sum_{mm'} (
C^{LM}_{lml'm'} )^2=1$.   Note that the absence of any $M$
dependence of  $\Xi^{LM}_{\rmA\rmA'}$ is as we expected.
Moreover, it follows from Eq.~(\ref{E:CGortho}) that the
estimators for the different $b_{lm}$s are uncorrelated: $\VEV{
\widehat b_{LM} \widehat b_{L'M'}} \propto \delta_{LL'}
\delta_{MM'}$.

Given a power spectrum of the form
Eq.~(\ref{E:powerspectrum}), specified by the functions
$g_{LM}(k)$, predictions for the $b_{LM}^{\rm
map}$ can be evaluated with Eq.~(\ref{E:blmtot}) replacing
$D^{LM}_{ll'}$ in that equation by $D^{LM,{\rm map}}_{ll'} =
D^{LM}_{ll'} W_l W_{l'}$ and evaluating the $D^{LM}_{ll'}$ with
Eq.~(\ref{E:cldl}).  The $b_{LM}^{\rm map}$ can then be measured
using Eq.~(\ref{E:dest}) with variances given by Eq.~(\ref{E:xlm}).

\subsection{A worked example} \label{S:limit}

As a simple example, suppose the $g_{LM}(k)$ are constants,
independent of $k$.  We can then take $g_{LM}$ outside the
integral in Eq.~(\ref{E:cldl}).  An estimator for $g_{LM}$ is
then $\widehat g_{LM} = \widehat b_{lm}^{\rm map}/(b^{\rm
map}_{lm}/g_{LM})$.  Defining $F_{ll'}\equiv
D^{LM}_{ll'}/g_{LM}$ for this case, the variance with which each
$g_{LM}$ can be measured is then
\begin{equation} \label{E:blmresult}
     \sigma_{g_{LM}}^2 = \frac{ 2 \sum_{ll'} G^L_{ll'}
     C_l^{\rm map} C_{l'}^{\rm map}} { \left[ \sum_{ll'} G^L_{ll'}
     F_{ll'} W_l W_{l'} \right]^2}.
\end{equation}
Moreover, the measured $g_{LM}$ are statistically independent as
a consequence of the statistical independence of the
$\widehat b_{LM}$.

To illustrate, we apply this result to an inflationary model
\cite{Ackerman:2007nb} that has a power spectrum with a
quadrupole dependence on the angle.\footnote{Note that our $g_{20}$ is
$(2/3)\sqrt{4\pi/5} g_*$, where $g_*$ is the coefficient in
Ref.~\cite{Ackerman:2007nb} of
$(\mathbf{\hat{k}}\cdot\mathbf{\hat{z}})^2$ if the preferred
direction is taken to be $\mathbf{\hat z}$.}
We use the $\Theta_l(k)$ calculated by CMBFAST \cite{cmbfast} to
obtain $F_{ll^\prime}$.  We assume only scalar perturbations and
the current best-fit cosmological parameters.

The numerical results are given in Table 1, but before reviewing
them, we provide some very rough estimates to get some feel for
the numbers.  To do so, ignore instrumental noise and suppose
that $W_l=1$ for all $l\leq l_{\rm max}$.  For $L=2$,
$F_{ll'}\neq0$ only for $l'=l$ or $l'=l\pm2$.  Moreover, for
these combinations of $ll'$ and for $l\gg2$, we approximate the
numerical results (which we use for the numerical results in the
Table) for $F_{ll'}$ as $F_{l,l+2}\simeq -0.5\,C_l$.
Also, $(C^{20}_{l0l0})^2\sim (5/8)l^{-1}$
for $l\gg1$, and $(C^{20}_{l0(l\pm2)0})^2$ is 1.5 times as
large.  Eq.~(\ref{E:blmresult}) can then be approximated
$\sigma_{g_{2M}}^2 \sim 256\pi [\sum_l l (C_l)^2] / [\sum_l l C_l]^2$.  
If the power spectrum has the
form $C_l \propto l^{-2}$ (a {\it very} rough approximation to
the temperature power spectrum for $l\lesssim1000$), then
$\sigma_{g_{2M}}^2 \sim 128\pi l_{\rm min}^{-2} [\ln(l_{\rm max}/l_{\rm
min})]^{-2}$.  For example, using $l_{\rm min}=2$ and $l_{\rm
max}=1000$ yields $\sigma_{g_{2M}} \sim 1.23$.  

Of course, there is nothing about the derivation of
Eq.~(\ref{E:blmresult}) that is specific to a temperature map,
and this result can be applied equally well, e.g., to the E-mode
polarization.  If we approximate the polarization power spectrum
by $C_l\sim{\rm const}$, then we find $\sigma_{g_{2M}}^2 \simeq
512\pi l_{\rm max}^{-2}$, or $\sigma_{g_{2M}} \sim 5\times
10^{-2}$ for $l_{\rm max}\simeq1000$.  

We now return to the numerical results for $\sigma_{g_{2M}}$
listed in Table 1 for the Wilkinson Anisotropy Probe (WMAP)
\cite{wmap}, which has now collected three years of data, the
Planck satellite \cite{planck}, to be launched in 2008, and EPIC
\cite{epic}, a satellite mission currently under study.  The
parameters assumed for each model are listed, as well as results
obtained using Eq.~(\ref{E:blmtot}) assuming only TT is used or
EE only.  The Appendix generalizes Eq.~(\ref{E:blmtot}) to the
case where the full temperature-polarization is used (including
the TE correlation), and we present numerical results for this
case in the Table as well.  We also list results, labeled
``CVO'' (cosmic variance only), for a hypothetical experiment that has
perfect angular resolution and no instrumental noise.  
These numbers are for hypothetical full-sky
experiments, but a realistic experiment will likely only be able to
use $\sim65\%$ of the sky for cosmology.  If so, then each
estimate for $\sigma_{g_{2M}}$ must be increased by a factor
$(0.65)^{-1/2}$, about 25\%.  We also note that the theory
cannot specify the direction $\mathbf{\hat e}$ of the
quadrupole, and so a search for a quadrupole would require
evaluation of all five $g_{2M}$s.  A ``$3\sigma$'' detection
would thus require that the sum of the squares of the $g_{2M}$s need
to exceed $(3\sigma_{g_{2M}})^2$, which is independent of $M$.

The order of magnitude that we would expect for $\sigma_{g_{2M}}$ 
is $\sim N_{\rm pix}^{-1/2}$, where $N_{\rm pix}\sim
l_{\rm max}^2$ is the number of resolution elements on the sky,
comparable to the precision with which one can measure the variance
(the monopole) of the temperature-fluctuation amplitude.  
The numerical results listed in Table 1 for the error to
$g_{2M}$ obtained from the power quadrupole moment $\widehat
b_{2M}$ are not quite as good as this $N_{\rm pix}^{-1/2}$
expectation. The origin of this discrepancy
can be traced to two sources.  First of all, the two-dimensional
CMB signal is degraded from the three-dimensional power
spectrum; a Fourier mode in the $\mathbf{\hat z}$ direction
gives rise to some temperature fluctuation near the north pole,
and not just at the equator.  This is manifest in the large
coefficients (e.g., the factor of $512\pi$) in our analytic
estimates.  

However, another reason that the estimator $\widehat
b_{2M}$ does not provide a sensitive probe of a quadrupole
departure from statistical isotropy is that it is not an optimal
estimator for $g_{2M}$.  This
estimator sums the ``signals'' $D^{LM}_{ll'}$, but it does not
weight these signals properly.  This can be seen by noting that
for a $C_l\propto l^{-2}$ power spectrum, for example, the error
obtained from Eq.~(\ref{E:blmresult}) can be reduced by applying
a low-pass filter: i.e., by increasing the minimum values of
$ll'$ in the sums.  (A simple calculation shows that with the
properly chosen lower-$l$ limit, $\sigma_{g_{2M}}$ can be
reduced by a factor of 30.)  If the precision of the result is improved
by removing data, then something is sub-optimal.  

\begin{table*}[tbp]
\begin{center}
\begin{tabular}{|l|c|c|c|c|c|c|c|c|c|}
\hline
Experiment & $\sigma_T~(\mu {\rm K})$ & $\sigma_P~(\mu {\rm K})$
& $\theta_{\rm fwhm}$ &  $\sigma_{g_{2M}}^{\rm pmm}$ (TT) &
$\sigma_{g_{2M}}^{\rm pmm}$ (EE) & $\sigma_{g_{2M}}^{\rm pmm}$ (total) &
$\sigma_{g_{2M}}^{\rm mv}$ (TT) & $\sigma_{g_{2M}}^{\rm mv}$
(EE) & $\sigma_{g_{2M}}^{\rm mv}$ (total) \\
\hline
\hline
WMAP & 30.0 &42.6 &21$^\prime$ & 1.3 & 11 & 1.2 & 0.024 & 2.4
& 0.024 \\
\hline
Planck &13.1 &26.8 &5$^\prime$ & 1.6 &0.16 &0.16 & 0.0052 &
0.033 & 0.0050 \\
\hline
EPIC &0.021 &0.068 &52$^\prime$ & 1.2 & 0.55 & 0.42 & 0.016 &
0.019 & 0.011 \\
\hline
Cosmic variance &0 &0 &0 & 1.8 &0.014 &0.014 & & & \\  
\hline  
\end{tabular}
\caption{The standard error $\sigma_{g_{2M}}$ to the amplitude of a
     quadrupole anisotropy in the matter power spectrum for
     different experiments.  The instrumental temperature and
     polarization noises and beam width are listed for each
     experiment.  We show results for the power multipole
     moments (pmm)for TT only, EE only, and the full result.  We also
     show in the last three columns $\sigma_{g_{2M}}^{\rm mv}$ from
     the minimum-variance estimator for each experiment, for TT
     only, EE only, and the full result.}
\end{center}
\end{table*}

\section{The minimum-variance estimator}
\label{S:optimal}

\subsection{The estimator and its variance}

The $\widehat b_{LM}$ estimator is a simple and intuitive
quantity that can be measured to test for statistical isotropy
in a model-independent way.  However, if one has a specific
theory, defined by the functions $g_{LM}(k)$ or some quantities
that parametrize the $g_{LM}(k)$, then there will be
estimators that can be constructed to measure optimally those
parameters.  For example, if the $g_{LM}$s are all constants,
then one can measure them better than the numerical results for
the power multipole moments $b_{LM}$ would suggest.  Below, we
will derive the minimum-variance estimator for $g_{LM}$.  

Before moving on, it is instructive and will be useful below to
re-derive the variance to $\widehat b_{LM}^{\rm map}$.  We return to
Eq.~(\ref{E:blmtot}) and note that $b_{LM}^{\rm map}$ can be written as a
sum over $D^{LM,\rm map}_{ll'}\equiv D^{LM}_{ll'} W_l
W_{l'}$.  We then return to Eq.~(\ref{E:cov}) to derive the
minimum-variance estimator for $D^{LM,\rm map}_{ll'}$.  Given a map
$a_{lm}^{\rm map}$, each $mm'$ pair provides an estimator
for $D^{LM,\rm map}_{ll'}$, through
\begin{equation} \label{E:Dllmmestimator}
     \widehat D^{LM,\rm map}_{ll',mm'} = \frac{ a_{lm}^{\rm map}
     a_{l'm'}^{{\rm map},*} - C_l \delta_{ll'}
     \delta_{mm'}}{\xi^{LM}_{lml'm'}},
\end{equation}
with variance
\begin{equation}
     \VEV{\left(\widehat D^{LM,\rm map}_{ll',mm'} \right)^2} =
     \frac{ (1+\delta_{ll'}\delta_{mm'}) C_l^{\rm map}
     C_{l'}^{\rm map}}{ \left( \xi^{LM}_{lml'm'}\right)^2}.
\end{equation}
The estimators for different $mm'$ pairs are uncorrelated (if
we use the real and imaginary parts of the $a_{lm}$s), so the
estimators can be summed over all $mm'$ pairs, inversely
weighted by the variance, to obtain a minimum-variance
estimator.  If $l=l'$, we sum only over $m'\geq m$ to avoid
double-counting pairs.  However, the factor $(1+\delta_{ll'}
\delta_{mm'})$ then weights the $m=m'$ modes twice as much, if
$l=l'$, and thus allows us to re-write the sum over all $m$ and
$m'$.  The result for the estimator can thus be written, for both
$l=l'$ and $l\neq l'$, as
\begin{equation} \label{E:DLMestimator}
     \widehat D^{LM,\rm map}_{ll'} = \frac{
     \sum_{mm'} a^{\rm map}_{lm} a^{{\rm map},*}_{l'm'}
     \xi^{LM}_{lml'm'} }{G^L_{ll'}}.
     \end{equation}
We recognize these to be the bipolar-spherical-harmonic
coefficients of Refs.~\cite{Hajian:2003qq}, with a slightly
different weight.  The variance of this estimator is then
\begin{equation}  \label{E:DLMllvariance}
     \VEV{ \left( \widehat D^{LM,\rm map}_{ll'} \right)^2 } =
     \frac{ (1+\delta_{ll'}) C_l^{\rm map}
     C_{l'}^{\rm map}}{ G^L_{ll'}}.
\end{equation}
The variance, Eq.~(\ref{E:blmresult}), with which each $b_{lm}$
can be measured simply follows by summing the variances of each
term in Eq.~(\ref{E:blmtot}).

Now, to construct the minimum-variance estimator, we simply note
that the statistically-independent quantities predicted by the theory
are the $D^{LM}_{ll'}$'s, the generalizations of the $C_l$'s for a
theory without statistical isotropy.  We have constructed above
estimators for these quantities, and we have their variances.
For a theory with constant $g_{LM}$'s, each $D^{LM}_{ll'}$
provides an estimator through $\widehat g_{LM,ll'} \equiv
\widehat D^{LM}_{ll'}/F_{ll'}$.  We then sum these,
inversely weighted by their variance to obtain the
minimum-variance estimator,
\begin{equation}
     \widehat g_{LM} = \frac{\sum_{l'\geq l} F_{ll'} W_l W_{l'}
     \widehat D^{LM,\rm
     map}_{ll'} \VEV{ \left( \widehat D^{LM,\rm map}_{ll'}
     \right)^2 }^{-1} }{\sum_{l'\geq l}  \left(F_{ll'} W_l
     W_{l'} \right)^2 \VEV{\left( \widehat D^{LM,\rm map}_{ll'}
     \right)^2 }^{-1}},
\end{equation}
obtained from the entire map.  The variance $\sigma_{g_{LM}}^2$
of this estimator is then obtained by summing the inverse
variances of all the estimators.  Again, the sums are over $l'\geq
l$, but the factor $(1+\delta_{ll'})$ in
Eq.~(\ref{E:DLMllvariance}) allows us to write the sum over all
$ll'$,
\begin{equation} \label{E:minvariance}
     \frac{1}{\sigma_{g_{LM}}^2} = \sum_{ll'} G^L_{ll'} \frac{
     (F_{ll'} W_l W_{l'})^2} { 2 C_{l}^{\rm map} C_{l'}^{\rm map}}.
\end{equation}

\subsection{Illustration: The Power Quadrupole}

To illustrate, we now evaluate this expression for $L=2$.
Again, in this case, the only $ll'$ combinations that
contribute are $l'=l$ and $l'=l\pm2$.  We assume
$l,l'\gg 1$, approximate $F_{l,l+2}\simeq -0.5\,C_l$, as above, and evaluate
$C^{LM}_{l0l'0}$ as in Section \ref{S:limit}. We can then write,
\begin{equation}  \label{E:minvarapprox}
     \frac{1}{\sigma_{g_{2M}}^2} \simeq 0.035\sum_l
     \frac{l C_l^2 (W_l)^4}{(C_l^{\rm map})^2},
\end{equation}
which we can further approximate as $0.017\, l_{\rm max}^2$, where
$l_{\rm max}$ is the multipole moment at which
$C_l^{\rmn} \simeq C_l (W_l)^2$.  The end result is then
$\sigma_{g_{2M}} \simeq 7.6/l_{\rm max}$, quite close to what we
would have expected by simply counting the number $N_{\rm pix}
\simeq l_{\rm max}^2$ of usable pixels.  For the WMAP and Planck
temperature maps, $l_{\rm max}$ is roughly 650 and 2000,
respectively, implying $\sigma_{g_{2M}}\sim 1.2\times10^{-2}$ and
$3.8\times 10^{-3}$, respectively, implying very significant
improvements in the sensitivity over the power multipole
moments.  

Table 1 lists the exact numerical results, obtained by
evaluating Eq.~(\ref{E:minvariance}) exactly, for both TT
only and EE only.  Again, the Appendix generalizes
Eq.~(\ref{E:minvariance}) for the full temperature-polarization
map, including the TE cross-correlation, and numerical results
for this case are also included.  The Table shows that 
by weighting the modes correctly, we get an improvement of a
factor of $\sim2$ for WMAP and Planck EE and more than an
order-of-magnitude improvement for WMAP and Planck TT; this is
in accord with our arguments that the
signal-to-noise in the TT power multipole moments was
particularly poorly chosen.  Although EPIC will have vastly
improved instrumental
sensitivity, with its modest angular resolution, it is not
particularly well suited to search for departures from
statistical isotropy.  Again, the minimum-variance numbers in
the Table must be increased by about 25\% to account for
partial-sky coverage.  And again, since the preferred direction is
not known {\it a priori}, the sum of the squares of the $g_{2M}$s must
exceed $(3\sigma_{g_{2M}})^2$ to claim a ``$3\sigma$'' detection of a
departure of statistical isotropy.

\section{Concluding Remarks} \label{S:conclusions}

We have considered CMB tests for the statistical isotropy of the
primordial power spectrum.  The power spectrum of
Eq.~(\ref{E:powerspectrum}) is the most general power spectrum
if the assumption of statistical isotropy is dropped.  In the
more general case, the CMB power spectrum $C_l$ is generalized
to a set of moments $D^{LM}_{ll'}$, which are closely analogous
to the bipolar-spherical-harmonic coefficients of
Refs.~\cite{Hajian:2003qq}.  The power multipole
moments $b_{LM}$ provide simple and intuitive statistics that
can be used to search in a model-independent way for
departures from statistical isotropy.  If, however, a particular
model is introduced by specifying a particular parametrization of
the functions $g_{LM}(k)$, then minimum-variance statistics
can be introduced to improve the precision with which these
parameters can be constrained.  For example, we constructed
explicitly the minimum-variance estimators for the coefficients
$g_{LM}$ for the case in which they are $k$-independent.  We
applied these results to a model in which there is a quadrupole
in the primordial power spectrum, and the results are shown in
Table 1.  We see that the best probe of a primordial quadrupole
moment will come from Planck TT, for which we anticipate
$\sigma_{g_{2M}}=0.0052$.  Multiplying this by 1.25 to account
for a 65\% sky coverage, and then by the factor of 3 required
for a  ``$3\sigma$'' detection, we find that the smallest
quadrupole amplitude that will be detectable by Planck will be
around 2.0\%.

To reduce clutter in the equations and to keep the main line of
reasoning clear, we 
have derived equations in the main body of the paper for the
case where either the temperature or the polarization is used,
but not both.  The Appendix generalizes the analysis to allow
the use of the full temperature-polarization information,
including the TE cross-correlation.

What about other probes?  Consider, for example, the Sloan
Digital Sky Survey \cite{sloan}.  The volume and galaxy density of the main
galaxy survey allows measurement, roughly speaking, of
the amplitudes of $N_{\rm modes}\sim10^5$ independent
Fourier modes of the density field, in the linear regime, and
these measurements are cosmic-variance limited.  Measurement of
the quadrupole of the power spectrum can then simply be done by
comparing the amplitudes of Fourier modes in different
directions.  The standard error to the power multipole moments
will thus be $\sigma_{g_{LM}}\sim \sqrt{2/N_{\rm modes}} \sim 10^{-2}$,
comparable in order of magnitude to what can be achieved with
the CMB.  Of course, a realistic search will be hampered by the
irregular volume of the survey, redshift-space distortions, and
anisotropies (line-of-sight--versus--angular) inherent to the
measurement technique.  But then again, there will be
degradations (foregrounds, sky cuts, etc.) to the idealized CMB
measurements we have considered.  Of course, if $g_{LM}(k)$ varies
with $k$, then the constraints provided by the CMB and galaxy
surveys will be complementary, to the extent that the
wavenumbers $k$ probed by the CMB and galaxy surveys differ.
Looking forward, there is ultimately the possibility of
accessing with 21-cm fluctuations approximately $10^{15}$ modes
of the primordial density field \cite{Loeb:2003ya}, allowing
values as small as $g_{LM}\sim 10^{-7}$ to be probed, but this
is in the very far future.

\acknowledgments

We thank D.~Babich, K.~Gorski, M.~Wise, and C. Pahud for useful comments
on an earlier draft.  MK acknowledges the hospitality of the
Aspen Center for Physics, where part of this work was completed.
AP acknowledges the support of the NSF.
This work was supported by DOE
DE-FG03-92-ER40701, NASA NNG05GF69G, the Gordon and Betty
Moore Foundation, and a NASA Einstein Probe mission study grant,
``The Experimental Probe of Inflationary Cosmology.''

\appendix

\section{Generalization to a Temperature-Polarization Map}

For most experiments, the sensitivity to departures from
statistical isotropy will come primarily from either the
temperature or the polarization.  Considering both in tandem
will provide some improvement in the result, but given the
temperature-polarization cross-correlation, this improvement
will be weaker than what would be obtained by simply adding the
two results in quadrature.

Still, to be complete, we include expressions for
theory and estimators for a combined temperature-polarization
map.  Assuming only primordial density perturbations contribute to the
temperature-polarization map, a map of the sky will now provide
the E-mode polarization $E(\hatn)$, constructed in the usual
fashion \cite{Kamionkowski:1996ks,Zaldarriaga:1996xe} from the
measured Stokes parameters $Q(\hatn)$ and
$U(\hatn)$, in addition to the temperature $T(\hatn)$.  The map
can be written in terms of spherical-harmonic coefficients
$a_{lm}^\rmX$, for $\rmX=\{\rmT,\rmE\}$, and Eq.~(\ref{E:cov})
is generalized to
\begin{equation} \label{E:covgeneral}
     \VEV{ a_{lm}^\rmX a_{l^\prime m^\prime}^{\rmX',*} }
     = \delta_{ll^\prime}\delta_{mm^\prime} C_l^{\rmX\rmX'}
     + \sum_{LM} \xi^{LM}_{lml^\prime  m^\prime}
     D^{LM,\rmX\rmX'}_{ll^\prime}.
\end{equation}
The $C_l^{\rmX\rmX'}$s and $D^{LM,\rmX\rmX'}_{ll^\prime}$s are
obtained as in Eqs.~(\ref{E:cls}) and (\ref{E:cldl}) by
replacing the $\Theta_l(k)\Theta_{l'}(k)$ factors in the
integrands of those equations by $\Theta_l^\rmX(k)
\Theta_l^{\rmX'}(k)$, where these are obtained from
Eq.~(\ref{E:temp}) by replacing $T(\hatn)$ by $X(\hatn)$.  Note
that for TE and $l\neq l'$, $D^{LM,\rmX\rmX'}_{ll'}\neq
D^{LM,\rmX\rmX'}_{l'l}$.  This will affect the equations below
for the minimum-variance estimator.

We now have a set of three power multipole moments
$b_{LM}^{\rmX\rmX'}$, obtained from Eq.~(\ref{E:blm}) by
replacing $\VEV{T^2}$ by $\VEV{XX'}$, which is itself obtained
from Eq.~(\ref{E:cn}) by using $\VEV{ a_{lm}^\rmX a_{l^\prime
m^\prime}^{\rmX',*} }$ for the expectation value therein.  The
expression for the $b_{LM}^{\rmX\rmX'}$ is the same as
Eq.~(\ref{E:blmtot}) using $D^{LM,\rmX\rmX'}_{ll'}$ there.

The power-multipole-moment estimators $\widehat
b_{LM}^{\rmX\rmX',\rm map}$ are as in Eq.~(\ref{E:dest}) with
$[T^{\rm map}(\hatn)]^2$ replaced by $[X^{\rm map}(\hatn)
X^{'\rm map}(\hatn)]$.  Things get a bit trickier, though, when
we calculate the variances, as the estimators for different
$\rmX\rmX'$ will now be correlated, although still uncorrelated
for different $LM$.  The variance in
Eq.~(\ref{E:covmat2}) is now promoted to a $3\times3$ matrix
$\Xi_{\rmA\rmA'}^{LM}$, for $\{\rmA,\rmA'\}
=\{\rmT\rmT,\rmE\rmE,\rmT\rmE\}$.  For
$\{\rmA\,\rmA'\}=\rmX\rmX^\prime=\{\rmT\rmT,\rmE\rmE\}$,
$\Xi_{\rmA\rmA'}^{LM}$ is given by Eq.~(\ref{E:xlm}) with
$C_{l_1}^{\rm{map}} C_{l_2}^{\rm{map}}$ replaced by
$C_{l_1}^{\rmA,\rm{map}} C_{l_2}^{\rmA^\prime,\rm{map}}$.
For the diagonal TE-TE term,
\begin{equation} \label{E:dsqte}
     \Xi_{\rmT\rmE,\rmT\rmE}^{LM} =
     \sum_{ll'} G^L_{ll'} [C_{l}^{\rmT\rmT,\rm{map}}
     C_{l'}^{\rmE\rmE,\rm{map}}+
     C_{l}^{\rmT\rmE,\rm{map}}C_{l'}^{\rmT\rmE,\rm{map}}],
\end{equation}
and for the off-diagonal XX-XX$'$ terms,
\begin{equation} \label{E:covarttte2}
     \Xi_{\rmX\rmX,\rmX\rmX^\prime}^{LM} =
     2\sum_{ll'} G^L_{ll'} C_{l}^{ \rmX\rmX,\rm{map}}C_{l'}^{ \rmX
     \rmX^\prime,\rm{map}}.
\end{equation}
Eq.~(\ref{E:blmresult}) for the standard error with which a
constant $g_{LM}$ can be recovered with the power multipole
moments is then replaced by
\cite{Jungman:1995bz,Kamionkowski:1996ks,Zaldarriaga:1996xe}
\begin{equation} \label{E:glim}
     \frac{1}{\sigma_{g_{LM}}^2} =
     \sum_{\rmA\rmA^\prime}\frac{\partial
     b_{LM}^A}{\partial
     g_{LM}}[ (\Xi^{LM})^{-1}]_{\rmA\rmA^\prime} \frac{\partial
     b_{LM}^{\rmA^\prime\ast}}{\partial g_{LM}}.
\end{equation}
This is the equation used to obtain the ``total'' results listed
in Table 1 for the power multipole moment.

The minimum-variance estimator for $g_{LM}$ and its variance 
are similarly generalized.  The estimators $\widehat
D^{LM,\rmA,\rm map}_{ll'}$ are still uncorrelated for different
$ll'$ pairs and different $LM$, but they are now correlated for
different A.  The main
subtlety is that since $D^{LM,\rmT\rmE}_{ll'} \neq
D^{LM,\rmT\rmE}_{l'l}$, we must be careful to keep track of all
TE modes for $l\neq l'$.  This will require that we split the
sum in the generalization of Eq.~(\ref{E:minvariance}) into two
sums: the first over $l=l'$, and the second over $l' > l$.
(Actually, the sum can in fact be written over all $ll'$, but at
the cost of much uglier algebraic expressions.)

For $l'=l$, there are now three (TT, EE, and TE) estimators to
replace that in Eq.~(\ref{E:Dllmmestimator}), and for $l'>
l$, there are now four (TT, EE, TE, and ET) estimators
to replace that in Eq.~(\ref{E:Dllmmestimator}).  For all $ll'$,
the estimators are as in Eq.~(\ref{E:Dllmmestimator}), replacing
each $a_{lm}^{\rm map}$ and $C_l^{\rm map}$ by the appropriate
$a_{lm}^{\rmX,\rm map}$ and $C_l^{\rmX\rmX',\rm map}$, respectively.
The estimator for each $ll'$, obtained after summing over all
$mm'$, is the same as in Eq.~(\ref{E:DLMestimator}).
For $l=l'$, the variances $\VEV{ \left( \widehat
D^{LM,\rmA,\rm map}_{ll'} \right)^2 }$ are now promoted to
a $3\times3$ covariance matrix.  and for $l'> l$, they are
promoted to a $4\times4$ covariance matrix.  In both cases, the
covariance matrix can be written as
\begin{equation}
     {\cal C}^{ll'}_{\rmA\rmA'} \equiv \frac{ G^L_{ll'}}
     {(1+\delta_{ll'})} \VEV{ \widehat D^{LM,\rmA,\rm map}_{ll'}
     \widehat D^{LM,\rmA',\rm map}_{ll'}}.
\end{equation}
For any $ll'$ pair, the diagonal entries, for
$\rmA=\{\rmT\rmT,\rmE\rmE\}$, are ${\cal C}^{ll'}_{\rmA\rmA} =
C_l^{\rmA,\rm map} C_{l'}^{\rmA,\rm map}$, and the TT-EE
off-diagonal entry is ${\cal C}^{ll'}_{\rmT\rmT,\rmE\rmE} =
C_l^{\rmT\rmE,\rm map} C_{l'}^{\rmT\rmE,\rm map}$. For $l=l'$,
the diagonal TE-TE entry is
\begin{equation}
     {\cal C}^{ll'}_{\rmT\rmE,\rmT\rmE}
     = \left[ C_l^{\rmT\rmT,\rm map} C_l^{\rmE\rmE,\rm map} + \left(
     C_l^{\rmT\rmE,\rm map}\right)^2 \right]/2.
\end{equation}
For $l'>l$, we
have ${\cal C}^{ll'}_{\rmT\rmE,\rmT\rmE} = C_l^{\rmT\rmT}
C_{l'}^{\rmE\rmE}$, ${\cal C}^{ll'}_{\rmE\rmT,\rmE\rmT} = C_{l'}^{\rmT\rmT}
C_{l}^{\rmE\rmE}$, and ${\cal C}^{ll'}_{\rmT\rmE,\rmE\rmT} = C_{l}^{\rmT\rmE}
C_{l'}^{\rmT\rmE}$.  For any $ll'$, we have ${\cal
C}^{ll'}_{\rmT\rmT,\rmT\rmE} = C_l^{\rmT\rmT} C_{l'}^{\rmT\rmE}$
and ${\cal C}^{ll'}_{\rmE\rmE,\rmT\rmE} = C_{l'}^{\rmE\rmE}
C_{l}^{\rmT\rmE}$.  For $l'>l$, we also have ${\cal
C}^{ll'}_{\rmT\rmT,\rmE\rmT} =  C_{l'}^{\rmT\rmT} C_{l}^{\rmT\rmE}$
and ${\cal C}^{ll'}_{\rmE\rmE,\rmE\rmT} = C_{l}^{\rmE\rmE}
C_{l'}^{\rmT\rmE}$. 

The generalization of Eq.~(\ref{E:minvariance}) is then
\begin{eqnarray} \label{E:XXprimegLM}
     \frac{1}{\sigma_{g_{LM}}^2} &=& \frac{1}{2}\sum_{l} G^L_{ll}
     \sum_{AA'} C^{\rmA}_{l} C^{\rmA'}_{l}
     (W_l)^4 \left[ \left({\cal C}^{ll}\right)^{-1}
     \right]_{\rmA\rmA'} \nonumber  \\
     &  + & \sum_{l'>l} G^L_{ll'} \sum_{\rmA\rmA'} F^\rmA_{ll'}
     F^{\rmA'}_{ll'} (W_l W_{l'})^2 \left[ \left( {\cal
     C}^{ll'}\right)^{-1} \right]_{\rmA\rmA'}, \nonumber \\
\end{eqnarray}
where the matrix inversion is in the $3\times3$ AA$'$ space in
the first sum and in the $4\times4$ AA$'$ space in the second sum.
We use Eq.~(\ref{E:XXprimegLM}) to evaluate the standard errors
for the ``total'' minimum-variance estimators listed in Table 1.

\end{document}